\documentclass[
amsmath,amssymb,
aps,
prd, 
twocolumn, 
amsmath,  
nofootinbib,
superscriptaddress,
showkeys
]{revtex4-2}
\usepackage[utf8]{inputenc}
\usepackage{graphicx}% Include figure files
\usepackage[T1]{fontenc}
\usepackage{float}
\usepackage{color}
\usepackage{amsmath}
\usepackage{multirow}
\usepackage{hyperref}% add hypertext capabilities
\makeatletter\def\Hy@Warning#1{}\makeatother
\usepackage[normalem]{ulem}
\hypersetup{colorlinks=true, citecolor=blue}
\usepackage{booktabs}
\usepackage{siunitx}

\graphicspath{ {} }

\begin{document}

\title{
Learning about neutron star composition from the slope of the mass-radius diagram}

\author{Márcio Ferreira}
\email{marcio.ferreira@uc.pt}
\affiliation{CFisUC, 
	Department of Physics, University of Coimbra, P-3004 - 516  Coimbra, Portugal}
	
\author{Constança Providência}
\email{cp@uc.pt}
\affiliation{CFisUC, 
	Department of Physics, University of Coimbra, P-3004 - 516  Coimbra, Portugal}

\begin{abstract}
The slope of the neutron star mass-radius curve, $dM/dR$, is studied to understand the information it may carry about the composition of neutron stars, particularly with regard to the presence of non-nucleonic degrees of freedom. This study uses two large sets of relativistic mean-field equations of state with either nucleonic or nucleonic and hyperonic degrees of freedom, and imposes constraints obtained from GW170817 and the pulsars PSR J0030+0451 and PSR J0740+6620. It is shown that:
i) some mass-radius curves are characterized by a negative slope from one solar mass up to the maximum mass;
ii) other equations of state (EoS) have a positive slope for a given range of masses below the maximum star mass. Within the set of models considered, the first set includes only a very small number of hyperonic EoS: less than 0.5\% of the total number of hyperonic stars and approximately one third of the nucleonic EoS. We have also analyzed the sign of the slope for neutron star masses of 1.2, 1.4 and 1.8$M_\odot$.  Only approximately 1\% of hyperonic equations of state (EoS) predict a negative slope for 1.4$M_\odot$ stars, whereas over 90\% of nucleonic stars have a negative slope at this mass. Finally, almost all stars have a negative slope at 1.8$M_\odot$. A positive slope at 1.4$M_\odot$ may indicate the presence of non-nucleonic degrees of freedom within neutron stars. The nuclear matter property that distinguishes the different scenarios most clearly is the curvature of the symmetry energy. Nucleonic EoSs with a positive slope $dM/dR$ predict the highest values, which can exceed 100 MeV.

\end{abstract}
\date{\today}

\maketitle

%\tableofcontents

\section{\label{sec:introduction} Introduction}

Neutron stars (NSs) are among the densest known objects in the Universe, yet their internal composition remains uncertain. A key challenge in nuclear astrophysics is determining the equation of state (EoS) of neutron star dense and isospin asymmetric nuclear matter that governs NS properties like mass, radius, and tidal deformability. It is still unclear whether exotic particles such as hyperons or deconfined quarks appear at supranuclear densities. These additional degrees of freedom could significantly alter the EoS, making their presence or absence a central focus of current research.\\

Observations of massive NSs have imposed stringent constraints on the EoS, especially at intermediate to high baryonic densities: $1.908 \pm 0.016~M_{\odot}$ for PSR J1614-2230 \cite{Demorest2010,Fonseca2016,Arzoumanian2017}, $2.01 \pm 0.04~M_{\odot}$ for PSR J0348-0432 \cite{Antoniadis2013}, $2.08 \pm 0.07~M_{\odot}$ for PSR J0740+6620 \cite{Fonseca:2021wxt}, and $2.13 \pm 0.04~M_{\odot}$ for PSR J1810+1714 \cite{Romani:2021xmb}. The rise of multi-messenger astrophysics, combining gravitational waves (GWs), electromagnetic signals, and neutrino emissions, has significantly enhanced our understanding of NSs. The detection of GW170817 \cite{Abbott:2018wiz} and GW190425 \cite{Abbott:2020khf} by LIGO/Virgo have provided additional constraints on the high-density EoS. Recent observational advances, particularly from the NICER (Neutron Star Interior Composition Explorer) mission, have led to precise mass and radius inferences, such as for PSR J0030+0451 \cite{Riley_2019,Miller19} and PSR J0740+6620 \cite{Riley2021,Miller2021,Raaijmakers2021}.  These efforts are expected to continue with upcoming missions, such as eXTP \cite{eXTP,eXTP:2018anb}, STROBE-X \cite{STROBE-X}, and the Square Kilometre Array (SKA) \cite{SKA}.\\

The composition of the NS core remains uncertain, especially concerning the potential presence of exotic matter such as deconfined quarks, meson condensates, or strange baryons (hyperons) \cite{book.Glendenning}. The determination of the mass of the pulsar J1614-2230, close to two solar masses has questioned the existence of hyperons inside NSs \cite{Demorest2010}. Hyperons are theoretically expected to emerge at high baryon densities, driven by the rapid increase in nucleon chemical potential. However, their presence typically softens the EoS, which in turn lowers the maximum mass that NSs can attain --- often falling below the observational threshold of $2M_{\odot}$ \cite{Baldo:1999rq}. 
This discrepancy with the observed masses of heavy NSs has become known as the hyperon puzzle \cite{Vidana:2010ip,Bednarek:2011gd, Weissenborn:2011ut,Providencia:2012rx,Lopes:2013cpa,Lonardoni:2014bwa,Tolos:2017lgv}; for a comprehensive review, see \cite{Chatterjee:2015pua}.
Before the determination of the mass of the pulsar PSR J1614-2232 close to two solar masses, the authors of \cite{Lackey:2005tk} had already discussed the presence of hyperons inside NSs considering the existing  NS observations on masses and redshifts, in particular the pulsar PSR J0751-1807, and concluded that only the stiffest hyperonic EoS would be compatible with observations. A similar conclusion was drawn several years later, considering already two solar-mass NSs in \cite{Fortin:2014mya}.  Recently, within an ab-initio microscopic Brueckner–Hartree–Fock theory which  includes  the strange baryon sector and considering the chiral hyperon-nucleon interaction of the Jülich–Bonn group tuned to the femtoscopic measurement of $\Lambda p$ obtained by the ALICE collaboration, the authors of \cite{Vidana:2024ngv}  have concluded that if only two-body hyperon-nucleon and hyperon-hyperon interactions are introduced, two solar mass neutron stars do not include hyperons. However, note that the EoS obtained is compatible with the tidal deformability of GW170817. Although some ab-initio calculations including hyperons are still not able to describe two solar mass NS possibly due to lack of information on three-body forces,  in \cite{Lonardoni:2014bwa}, within the auxiliary field diffusion Monte Carlo algorithm, the authors found that three-body hyperon-nucleon interaction could explain the observation of very massive neutron stars.   Phenomenological descriptions of the NS EoS based on covariant density functionals constrained by hypernuclei properties have succeeded in describing two solar mass NS \cite{Bednarek:2011gd,Weissenborn:2011kb,Providencia:2012rx,Lopes:2013cpa,Tolos:2017lgv,Sun:2022yor}. As shown in Ref. \cite{Huang:2024rvj}, present observations of neutron star masses and radii, or tidal deformabilities and their associated uncertainties, are insufficient to identify hyperons within neutron stars. 
\\

The mass-radius relation \( M(R) \) of neutron stars encodes the properties of the underlying EoS of neutron star matter. Its slope, $dM/dR$, reflects the EoS stiffness: steeper slopes correspond to stiffer EoS. Microscopic and phenomenological models predict distinct $dM/dR$ behaviors \cite{Ferreira:2024hxc}, and transitions on this slope may signal the emergence of new degrees of freedom, such as hyperons or deconfined quarks, making changes in the internal composition of the star. In addition, the onset of a new degree of freedom together with the neutron star two solar mass  constraint may cause a stiffening of the  EoS below 2-3 times saturation density, even if the new degree of freedom is still not present, to compensate the softening at higher densities. This indicates that the analysis of low-mass stars may give information on the composition of the high mass neutron stars. In other studies, the physical meaning of the slope of the mass-radius curve has also been discussed, see \cite{Li:2024imk,Ayriyan:2024zfw,Cai:2025nxn}. In \cite{Tan:2021nat}, the authors have discussed how the slope of the binary Love relations reflects the rate of change of the nuclear matter speed of sound using an agnostic description of the EoS. It was also shown that a steep rise in the speed of sound at low (high) densities would give rise to a positive (negative) slope of the mass-radius curve. The effect of the mass-radius curve slope on NS properties, in particular on the  quasi-universal relations between the peak spectral frequency $f_2$ of the GW spectrum  and the stellar radius, was also discussed in \cite{Raithel:2022orm}. In particular,  using a piecewise polytropic parameterization of EoS, the authors have shown that backward bending slopes in mass–radius relations violate the universal relations proposed in \cite{Bauswein:2011tp}. 
It is, therefore, interesting to explore what kind of information the slope of the mass-radius curve may give when EoS based on a microphysics description are considered. Note that in both the studies \cite{Cavagnoli:2011ft,Dexheimer:2018dhb} it was shown that the mass-radius slope for low to medium mass stars may be influenced by the properties of the symmetry energy: if the isoscalar channel of the EoS is fixed and only the symmetry energy varied, EoS with a smaller symmetry energy slope at saturation give rise to   low  mass stars with smaller radii, and a backbend may appear. 
 \\

In this work, we explore the quantity $dM/dR$ as a diagnostic observable to distinguish between hadronic and hyperonic EoS. Using a family of relativistic mean-field (RMF) models constrained by empirical data and capable of producing NSs with $M \gtrsim 2M_{\odot}$, we study the behavior of $dM/dR$ across the mass-radius space. Our analysis emphasizes how the emergence of hyperons or changes in the stiffness of the EoS may influence this slope. By systematically comparing the hadronic and hyperonic model sets, we aim to quantify differences in $dM/dR$ and assess their observational signatures.
Note that the hyperonic EoS H4, proposed in \cite{Lackey:2005tk} within a covariant density functional description, has frequently been tested against results from GW170817, showing a very low compatibility with the data \cite{Abbott:2018wiz,Ghosh:2021eqv}. This is mainly due to the fact that it is a very stiff EoS which does not satisfy several nuclear matter properties at saturation, having a large incompressibility (300 MeV)  and a symmetry energy slope close to 100 MeV. However, within the same formalism, it is possible to build EoS including hyperons compatible with the GW170817 data \cite{Sun:2022yor,Malik:2022jqc,Malik:2023mnx,Kochankovski:2023trc}. \\

The paper is structured as follows: In Section \ref{sec:models}, we present the nucleonic and hyperonic datasets analyzed in the present work. In Section \ref{sec:results} we discuss how the value of $dM/dR$ determined from the solution of the TOV equations affects the thermodynamical properties, the astrophysics predictions and the nuclear matter properties within the  nucleonic and hyperonic scenarios. We access the likelihood of each composition scenario given the present observational data in Section \ref{sec:loglikelihood}.
Finally, conclusions are drawn in Section \ref{sec:conclusions}.

\section{\label{sec:models} Nucleonic and Hyperonic Neutron Star Matter}

We analyze two datasets introduced in \cite{Malik:2023mnx}, both constructed using a relativistic mean field (RMF) model with nonlinear meson interactions. The first dataset describes nuclear matter composed exclusively of nucleons, while the second includes additional degrees of freedom in the form of hyperons—specifically, the neutral $\Lambda$ hyperon and the negatively charged $\Xi^-$ hyperon.
These datasets consist of posterior samples obtained through a Bayesian inference framework that incorporates empirical constraints from nuclear matter saturation properties, low-density pure neutron matter, and $M_{\rm max}>2.0 M_{\odot}$. The nucleonic dataset contains 17,828 EoS, while the hyperonic dataset comprises 18,756 EoS.\\

We imposed the following astrophysical constraints, all given at the 95\% confidence level:
i) For the pulsar PSR J0740+6620, we adopted the radius intervals
$10.71\,\text{km} < R(2.07M_\odot) < 15.02\,\text{km}$ from \cite{2021ApJ...918L..27R} and
$11.14\,\text{km} < R(2.06M_\odot) < 20.20\,\text{km}$ from \cite{2021ApJ...918L..28M}.
ii) For PSR J0030+0451, we used
$10.94\,\text{km} < R(1.44M_\odot) < 15.50\,\text{km}$ from \cite{2019ApJ...887L..24M} and
$10.57\,\text{km} < R(1.34M_\odot) < 14.86\,\text{km}$ from \cite{2019ApJ...887L..21R}.
To incorporate the PSR J0740+6620 constraints into our analysis, we required that
$R(2.0M_\odot) > 10.71\,\text{km}$.
This condition is less restrictive than $R(2.07M_\odot) > 10.71\,\text{km}$, and accounts for the uncertainty in the mass measurement of PSR J0740+6620.
Additionally, we imposed a tidal deformability constraint of $\Lambda(1.4M_\odot) < 720$, following the results from \cite{Abbott:2018wiz}. After applying these constraints, the number of EoS was reduced to 17,537 for the nucleonic set and  
16,146 for the for the hyperonic one.\\

For each EoS, we calculated the slope $dM/dR$ along the Tolman–Oppenheimer–Volkoff (TOV) sequence in the mass range from $1.0M_{\odot}$ up to the maximum supported mass $M_{\mathrm{max}}$. This slope reflects how small variations in the radius of a neutron star affect its mass.
Based on this criterion, we found that out of 17,537 nucleonic EoS, 11,495 exhibit  $dM/dR < 0$, while the remaining 6,042 do not (i.e., $dM/dR \not < 0$).
In the hyperonic case, among 16,146 EoS, only 77 satisfy $dM/dR < 0$, whereas 16,069 fail to meet this condition (i.e., $dM/dR \not < 0$). The nucleonic and hyperonic EoS show a different behavior in terms of the mass dependence of $dM/dR$: most nucleonic EoS fulfill the condition $dM/dR < 0$, in contrast to the hyperonic EoS, where only a small fraction satisfies it. Note that in the present work, similar to \cite{Ferreira:2024hxc}, $dM/dR$ refers to a global property of the neutron star mass-radius sequence, and we use the following notation: $dM/dR < 0$ indicates that the radius decreases monotonically when mass increases  for all stable NSs;  $dM/dR \not< 0$ implies that this behavior is not monotonic, and when the mass increases in some range the radius decreases, in other the  radius increases or even keeps constant.

In the following, we compare both nucleonic/hyperonic sets and analyze the impact of $dM/dR < 0$ or $dM/dR \not< 0$ on the stars' properties and thermodynamics of neutron star matter. In Sec. \ref{sec:results2}, local quantities \( dM/dR|_M \) at specific NS masses are analyzed.

\section{\label{sec:results} Results}

In the present section we discuss the properties of the sets of nucleonic and hyperonic NS characterized by a negative or positive slope along the whole curve or at given NS masses.

\subsection{\label{sec:results1} General behavior of $dM/dR$}
\begin{figure*}[!htb]
    \centering
    \includegraphics[width=0.329\linewidth]{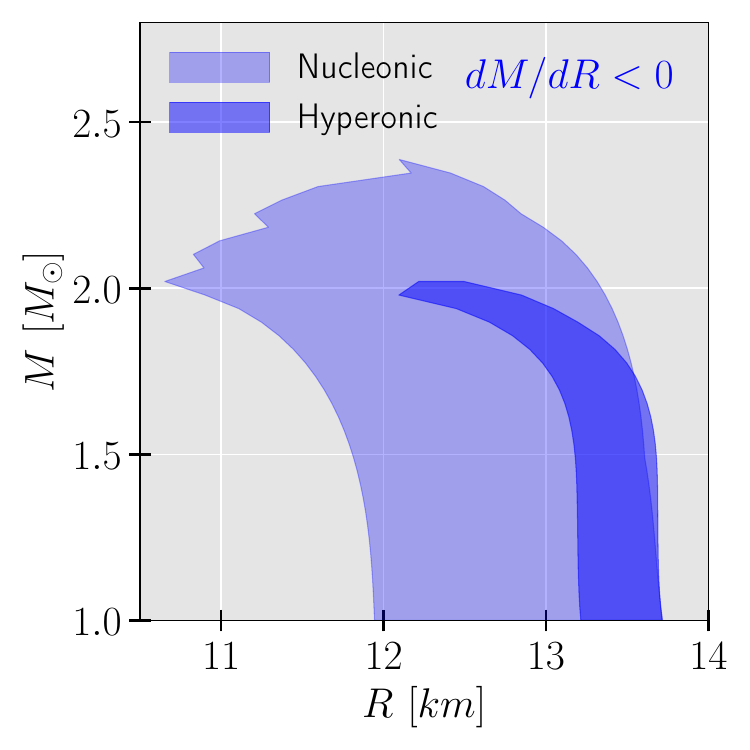}
    \includegraphics[width=0.329\linewidth]{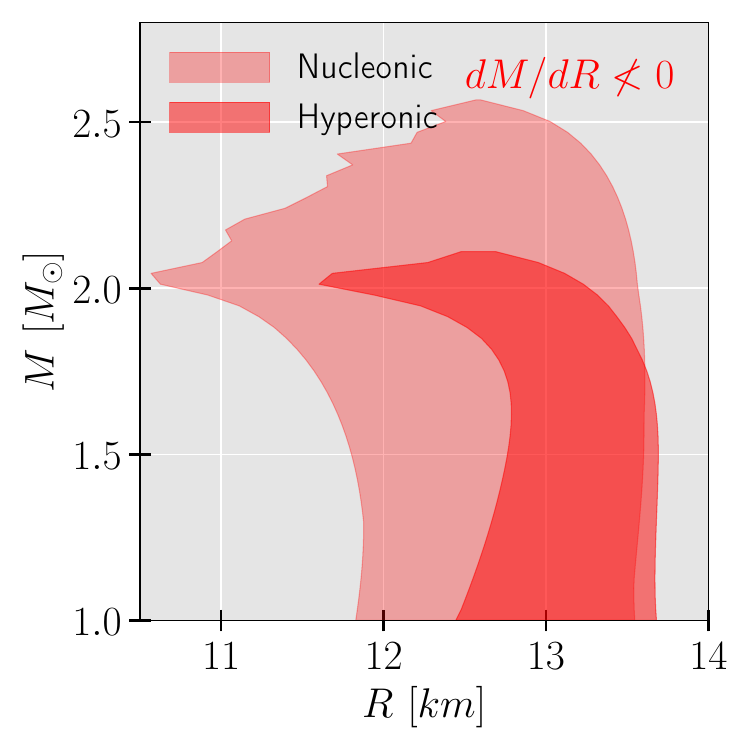}
    \includegraphics[width=0.329\linewidth]{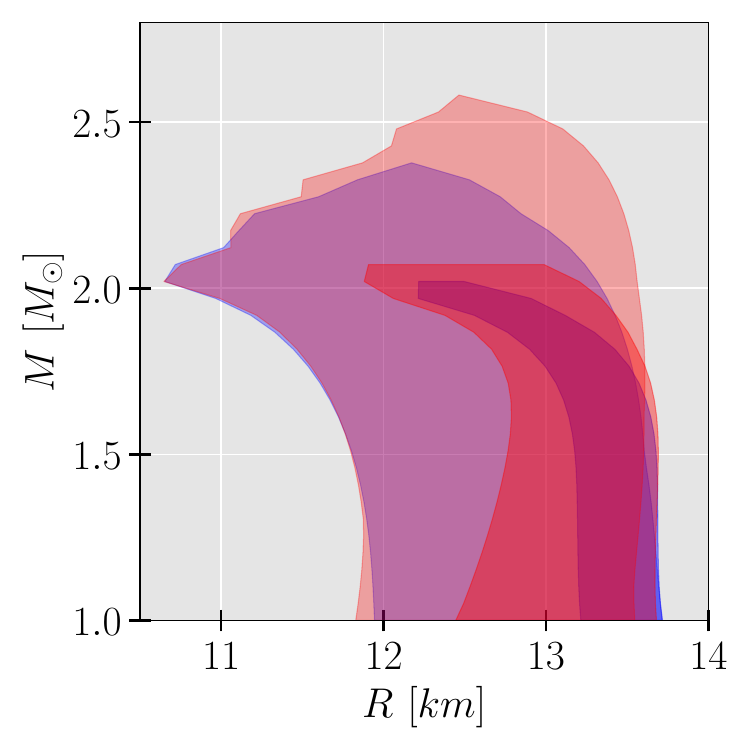}
    \caption{Mass-radius relations for different sets of models. {The full probability distributions have been plotted. To complement these plots the 90\% CI of several NS properties are given in Table \ref{tab:1}.} Colors indicate the slope of the mass-radius curve: blue for $dM/dR < 0$ and red for $dM/dR \not < 0$. Light shades represent nucleonic sets, while dark shades represent hyperonic sets. Left panel: Sets without back-bending ($dM/dR < 0$). Middle panel: Sets with back-bending ($dM/dR \not < 0$). Right panel: All sets combined. The total number of 17,537 nucleonic EoS are distributed by 11,495 exhibiting  $dM/dR < 0$ and 6,042 with $dM/dR \not < 0$. From the 16,146 EoS hyperonic EoS, only 77 satisfy $dM/dR < 0$ while 16,069 $dM/dR \not < 0$.}
    \label{fig:1}
\end{figure*}

The $M(R)$ relation for the nucleons and hyperons datasets, filtered by the condition $dM/dR < 0$ and $dM/dR \not < 0$, is shown in Fig. \ref{fig:1}.

{We first consider the nucleonic sets. From the general structure, the main difference is the set $dM/dR\not<0$ predicting larger radii for stars $M\gtrsim 1.4 M_\odot$ and the prediction of larger maximum masses well above $\sim 2.35 M_\odot$. The probability distributions of the two sets are essentially identical for the minimum allowed radius of all masses.

Comparing the $dM/dR<0$ and $dM/dR\not<0$ for both the nucleonic and the hyperonic sets,  we find that the hyperonic case has a much smaller range of possible radii, corresponding to one third for $dM/dR<0$ and one half for $dM/dR\not<0$ of the nucleonic case.  Besides the hyperonic distribution almost coincides with the nucleonic one at the maximum radius limit for masses below $\sim 1.6\, M_\odot$, and  predicts a smaller maximum mass of about 0.3$M_\odot$ for $dM/dR<0$   and $0.5M_\odot$ for $dM/dR\not<0$. Finally the hyperonic set also gives a slightly larger radius for $\sim 1.4M_\odot$ stars.}\\

{Table \ref{tab:1} provides complementary information to Fig.~\ref{fig:1}, showing the 5\%, 50\% (median), and 95\% quantiles of selected neutron star (NS) properties for the two datasets.} By comparing the radius predicted by the hyperon and nucleon sets for 1.4$M_\odot$ stars, we see that the hyperonic EoS do not predict radii below $\sim 13$ km, and that all radii for $M\sim 1.4M_\odot$ are within a boundary interval of $\sim 600$ m.  The set with a positive slope predicts a radius $\sim$150–300 m smaller. For nucleonic stars, this radius varies between $\sim$12 and 13 km; larger radii occur for curves with a positive slope. 
The hyperonic maximum mass is about $2M_\odot$ with a radius between 11.6 and 12 km. The two scenarios, positive and negative slope, give similar results. For the nucleonic set with $dM/dR<0$, maximum masses of up to $2.4M_\odot$ (at {90}\% CI) are possible, with radii ranging from 10.6 km to below 12 km. $2M_\odot$ stars have radii of $12\lesssim R\lesssim 13$ km. A positive slope allows for maximum nucleonic star masses $0.2M_\odot$ larger at {90}\% CI and radii up to 500 m larger.

\begin{figure*}[!htb]
    \centering
    \includegraphics[width=0.329\linewidth]{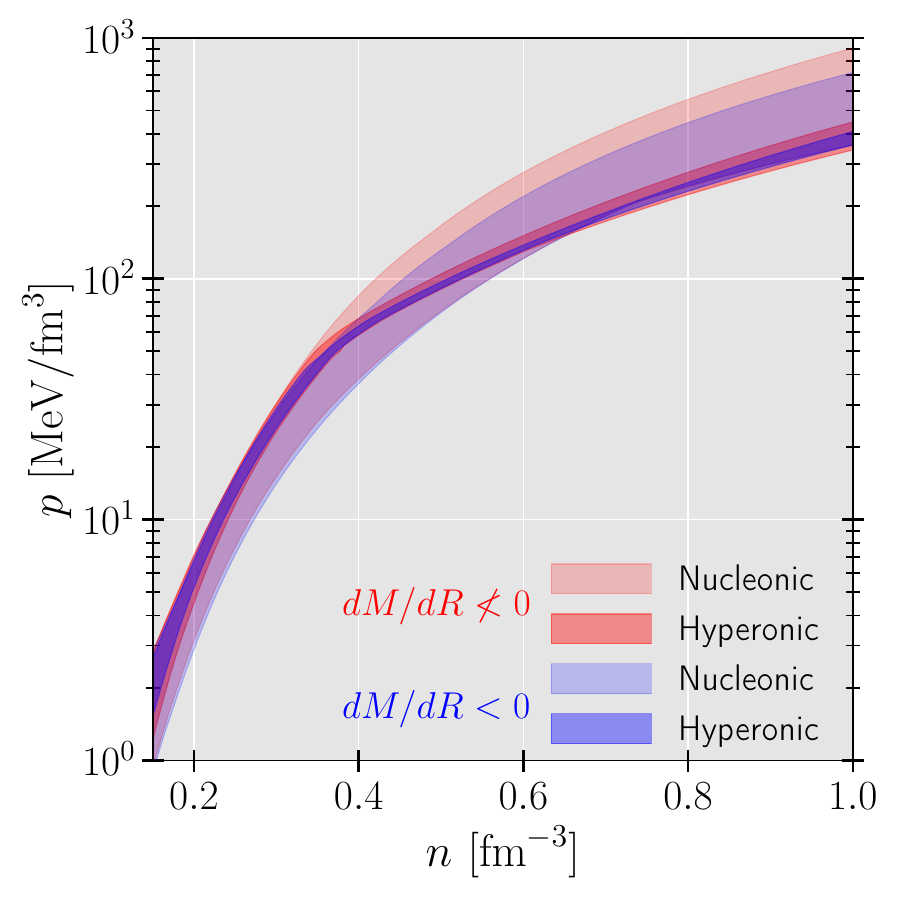}
    \includegraphics[width=0.329\linewidth]{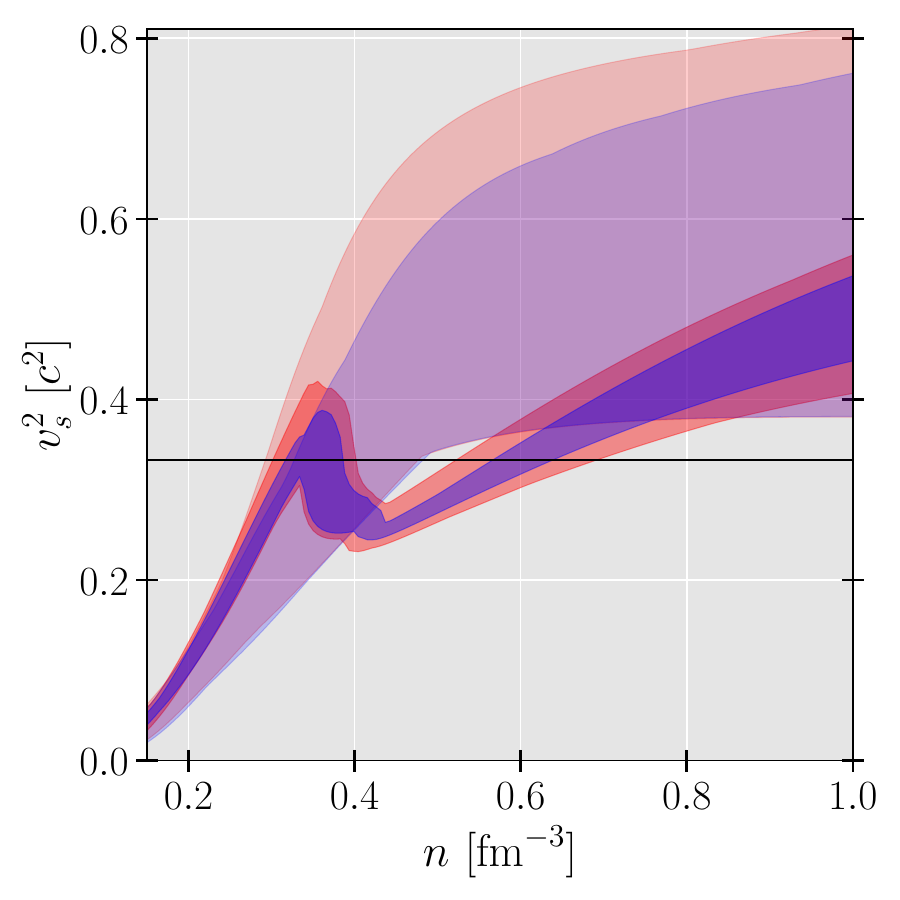}
    \includegraphics[width=0.329\linewidth]{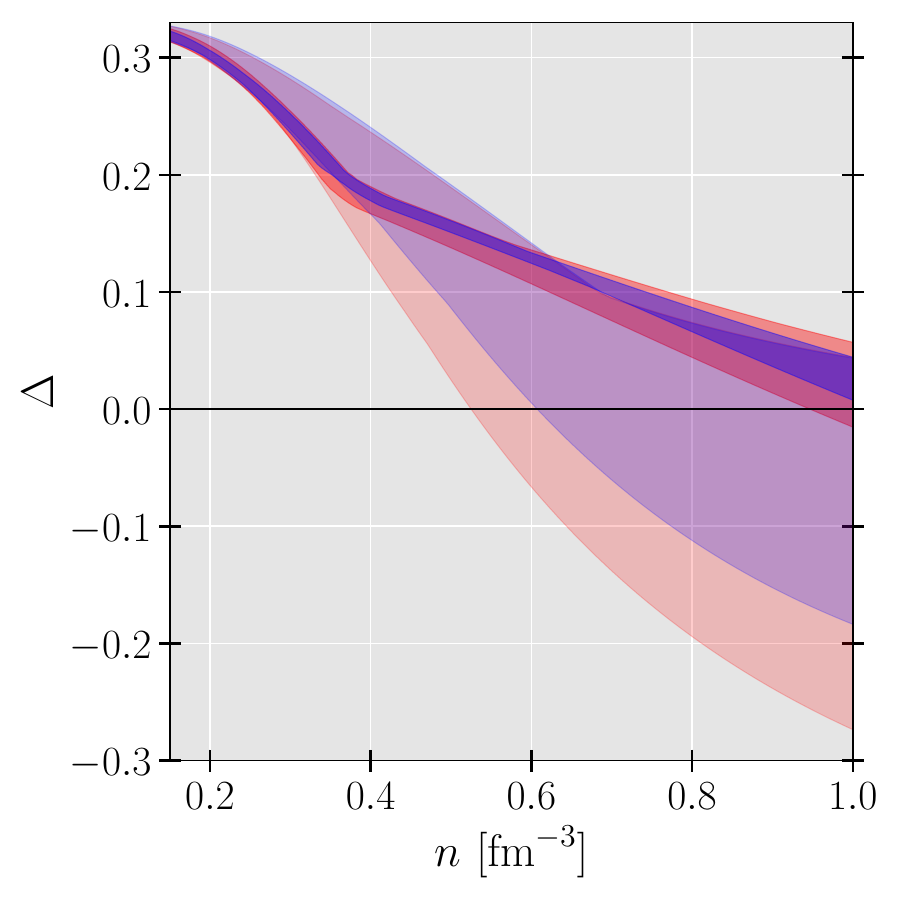}
    \caption{The pressure (left), speed-of-sound squared (middle), and  the renormalized trace anomaly (right) as a function of baryonic density for the four sets displayed in Fig.~\ref{fig:1}. {Full distributions are given.} Colors indicate the slope of the mass-radius curve: blue for $dM/dR < 0$ and red for $dM/dR \not < 0$. Light shades represent nucleonic sets, while dark shades represent hyperonic sets. }
    \label{fig:2}
\end{figure*}

Figure \ref{fig:2} displays the pressure $p(n)$, the squared sound speed $v_s^2(n)$ and the renormalized trace anomaly $\Delta(n) = 1/3 - p(n)/e(n)$, where $e(n)$ is the energy density. The trace anomaly $\Delta$ is a key measure for studying conformality restoration in neutron star matter \cite{Fujimoto:2022ohj}. For nucleonic EoS, the lower limit of $p(n)$ is the same for both sets ($dM/dR < 0$ and $dM/dR \not < 0$), since this defines the condition of a maximum mass of at least two solar masses. This also sets the lower limit for $v_s^2(n)$ and the highest positive values of $\Delta(n)$, this last quantity remaining positive throughout the entire density range.
When comparing hyperonic and nucleonic EoS, the hyperonic EoS tracks the upper boundary of the nucleonic EoS at densities below 2$n_0$ and the lower boundary above 4$n_0$. The hyperonic set $dM/dR < 0$ lies within the set $dM/dR \not < 0$. Below 2$n_0$, the $v_s^2(n)$ aligns with the upper limit of the nucleonic set, a condition necessary to compensate for the softening occurring above 2$n_0$. At 2$n_0$, $v_s^2$ decreases due to the appearance of hyperons, then increases, approaching the nucleonic lower limit at a density of 1 fm$^{-3}$. At this density, the trace anomaly $\Delta$ remains positive for the hyperonic set $dM/dR < 0$.

\begin{table*}[!htb]
\caption{The 5\%, 50\%, and 95\% quantiles for some NS and EoS properties {shown in Figs. \ref{fig:1} and \ref{fig:2}, in particular, the mass and radius of the maximum mass configuration together with its  baryonic density, speed of sound squared, pressure and trace anomaly at the center, as well as the radius and tidal deformability of a 1.4 $M_\odot$ star and the radius of a 2.0 $M_\odot$ star.}}
\label{tab:1}
\centering
%\resizebox{\columnwidth}{!}{%
\begin{tabular}{c ccccccc c ccccccc }
\toprule
&  \multicolumn{7}{c}{Nucleonic} & &  \multicolumn{7}{c}{Hyperonic} \\
\cline{2-8} \cline{10-16} 
& \multicolumn{3}{c}{$dM/dR<0$}  & &\multicolumn{3}{c}{$dM/dR\nless0$} & &\multicolumn{3}{c}{$dM/dR<0$}  & &\multicolumn{3}{c}{$dM/dR\nless0$}  \\
\cline{2-4} \cline{6-8} \cline{10-12} \cline{14-16} 
& 5\% & 50\% & 95\% & & 5\% & 50\% & 95\% & & 5\% & 50\% & 95\% & & 5\% & 50\% & 95\%\\ 
 \midrule
 $M_{\mathrm{max}}$ [$M_{\odot}$]  &2.01&2.05&2.17 & &2.01&2.11&2.35 & &2.00&2.01&2.04 & &2.00&2.02&2.07  \\
$R_{\mathrm{max}}$ [km]  &10.52&10.84&11.39 & &10.62&11.16&11.92 & &11.59&11.8&12.0 & &11.54&11.79&12.04  \\
$n_{\mathrm{max}}/n_0$ &6.05&6.67&7.05 & &5.39&6.25&6.88 & &5.66&5.89&6.07 & &5.53&5.80&6.06  \\
$v_s^2(n_{\mathrm{max}})$ [$c^2$] &0.46&0.58&0.68 & &0.43&0.56&0.69 & &0.44&0.47&0.52 & &0.42&0.47&0.51  \\
$p(n_{\mathrm{max}})$ [MeV/fm$^{3}$] &414.79&529.26&606.02 & &367.49&484.84&578.08 & &306.13&337.57&372.42 & &290.44&329.06&364.11  \\
$\Delta(n_{\mathrm{max}})$ &-0.100&-0.056&0.000 & &-0.115&-0.050&0.017 & &0.022&0.040&0.057 & &0.024&0.041&0.063  \\
$R(1.4M_{\odot})$ [km] &12.09&12.40&12.87 & &12.10&12.56&13.15 & &13.26&13.39&13.55 & &12.96&13.19&13.43  \\
$\Lambda(1.4M_{\odot})$  &344&406&513 & &376&476&632 & &592&640&683 & &593&650&709  \\
$R(2.0M_{\odot})$ [km] &10.86&11.45&12.27 & &11.08&11.97&12.96 & &11.83&12.15&12.66 & &11.84&12.33&12.85  \\
\bottomrule
\end{tabular}
%}
\end{table*}

\subsection{\label{sec:results2} Behavior of $dM/dR$ at specific masses}
We are interested in understanding whether an observational estimate of the derivative \( dM/dR|_M \) at specific neutron star (NS) masses could be used to constrain nuclear matter properties, with particular emphasis on isovector properties.  We consider both nucleonic and hyperonic matter scenarios. {The two  EoS datasets determined in  \cite{Malik:2023mnx}  were published in \cite{malik_2023_7854112} together with the NS and the saturation nuclear matter properties. This information will allow us to develop the analysis of the present subsection. } Given that accurately measuring the value of \( dM/dR|_M \) is observationally challenging, we focus primarily on determining its sign (whether it is positive or negative).

{We calculate the sign (positive or negative) of 
\( dM/dR|_M \) 
for all EoS at fixed stellar masses 
$M/M_{\odot} = 1.2, 1.4, 1.8$. 
Based on this classification by sign, we then plot the probability density functions (PDFs) of the isovector properties, separately for the nucleonic and hyperonic datasets, using kernel density estimation. 
} {The saturation nuclear matter properties of each set (nucleonic and hyperonic with  positive and negative \( dM/dR|_M \))  are summarized in Table \ref{tab:3}.}

\begin{figure*}[!htb]
    \centering
    \includegraphics[width=0.32\linewidth]{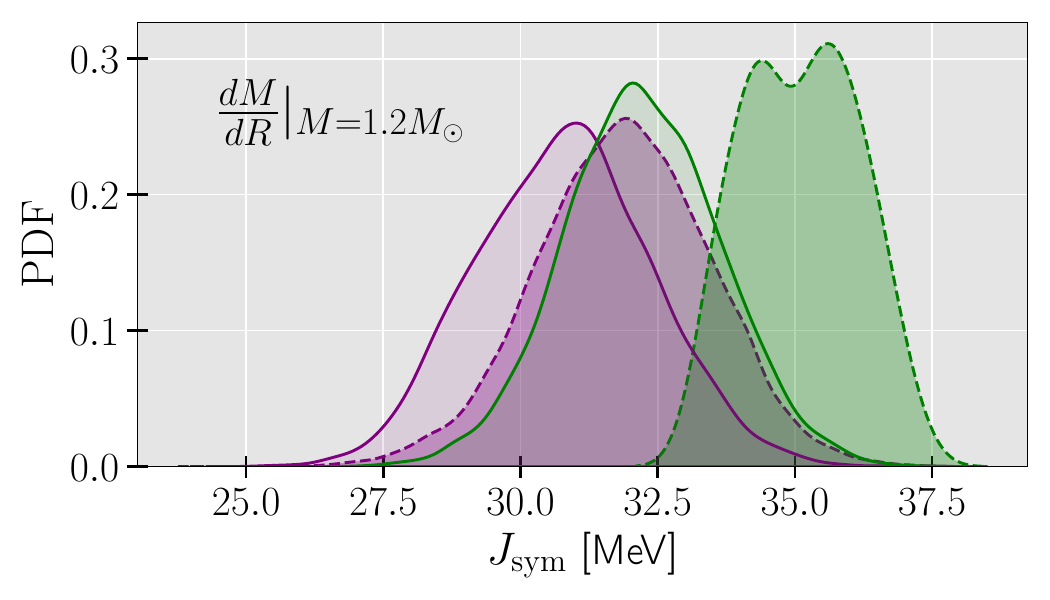}
    \includegraphics[width=0.32\linewidth]{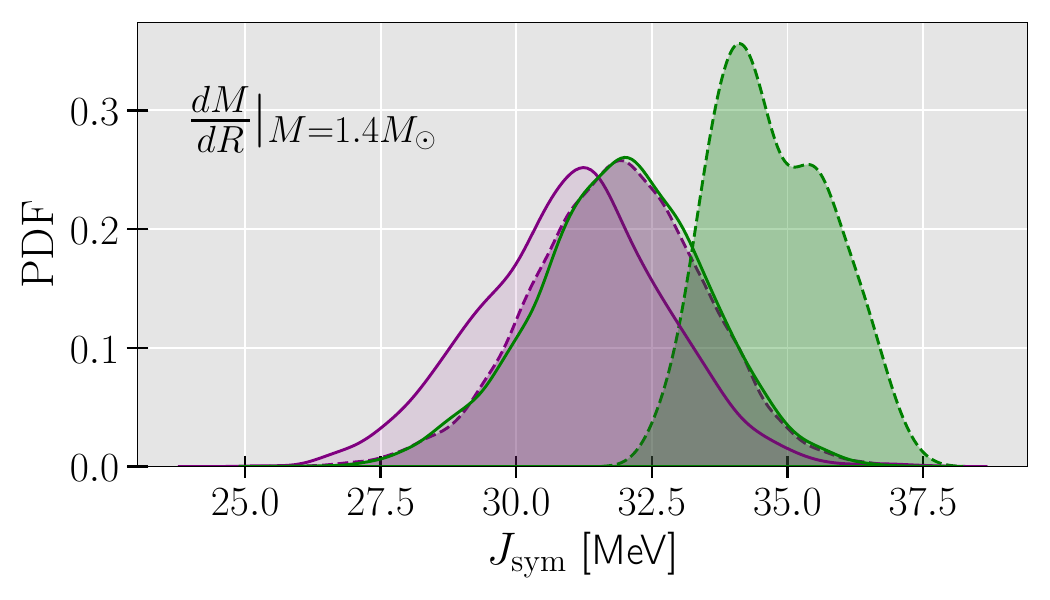}
    \includegraphics[width=0.32\linewidth]{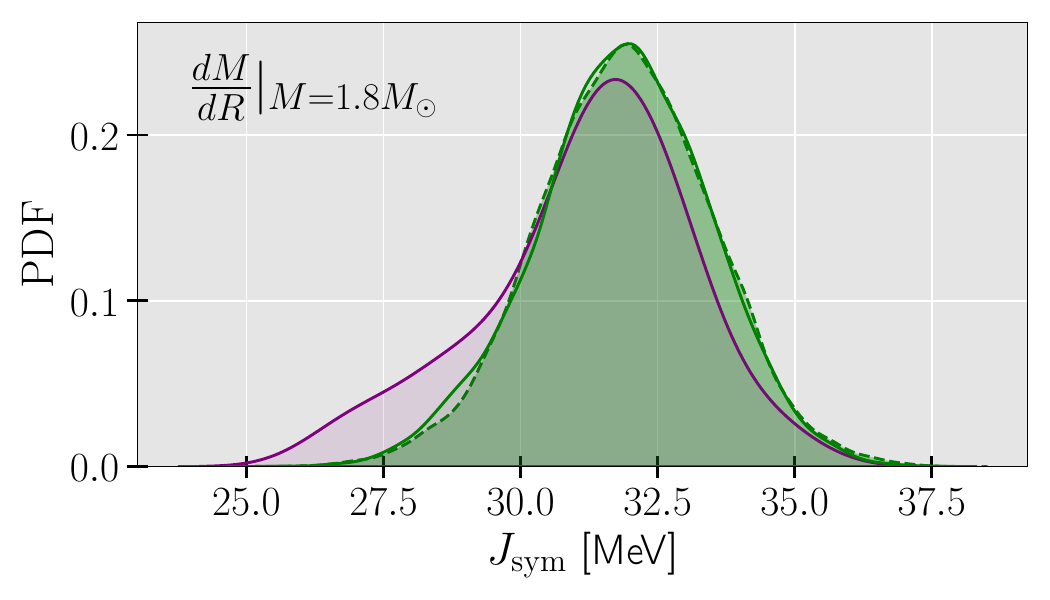}\\
    \includegraphics[width=0.32\linewidth]{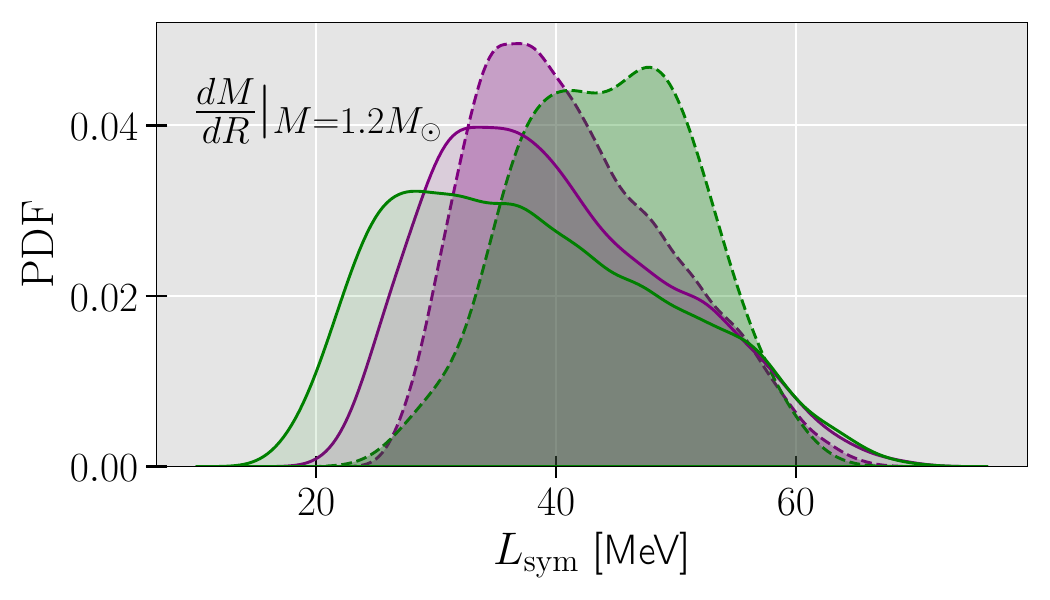}
    \includegraphics[width=0.32\linewidth]{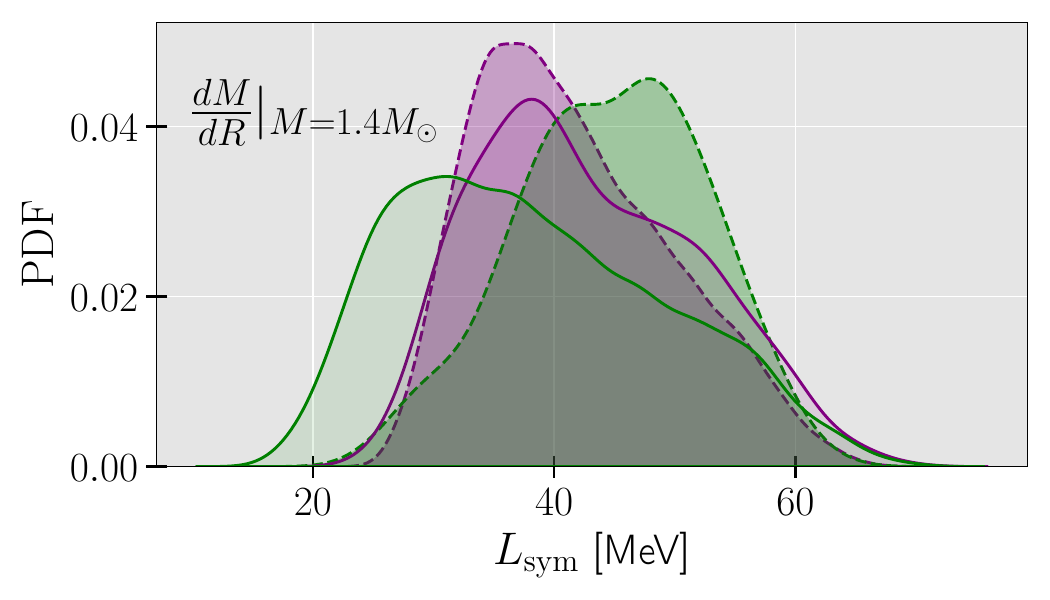}
    \includegraphics[width=0.32\linewidth]{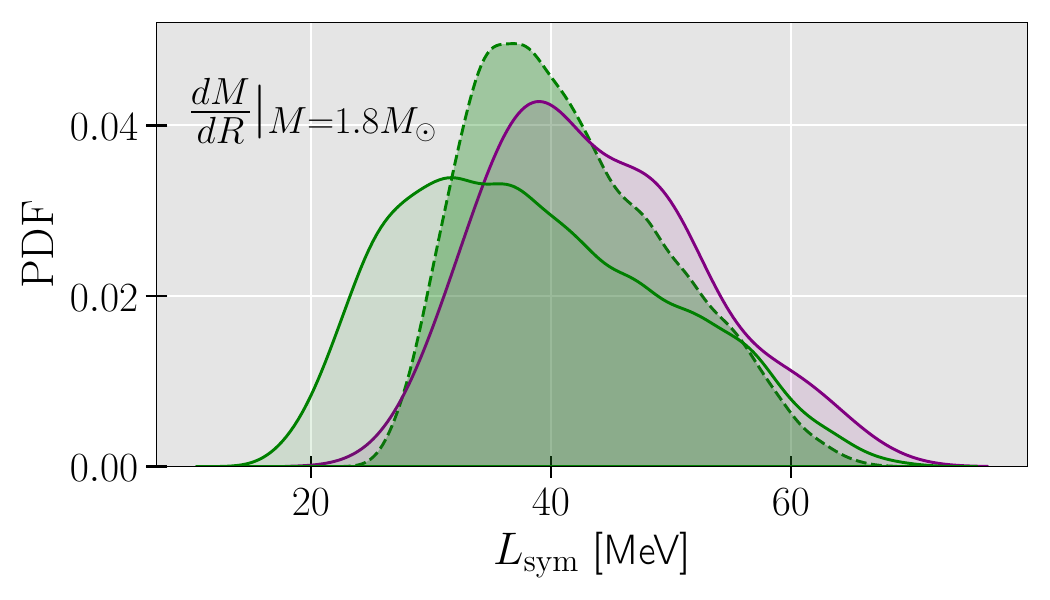}\\
    \includegraphics[width=0.32\linewidth]{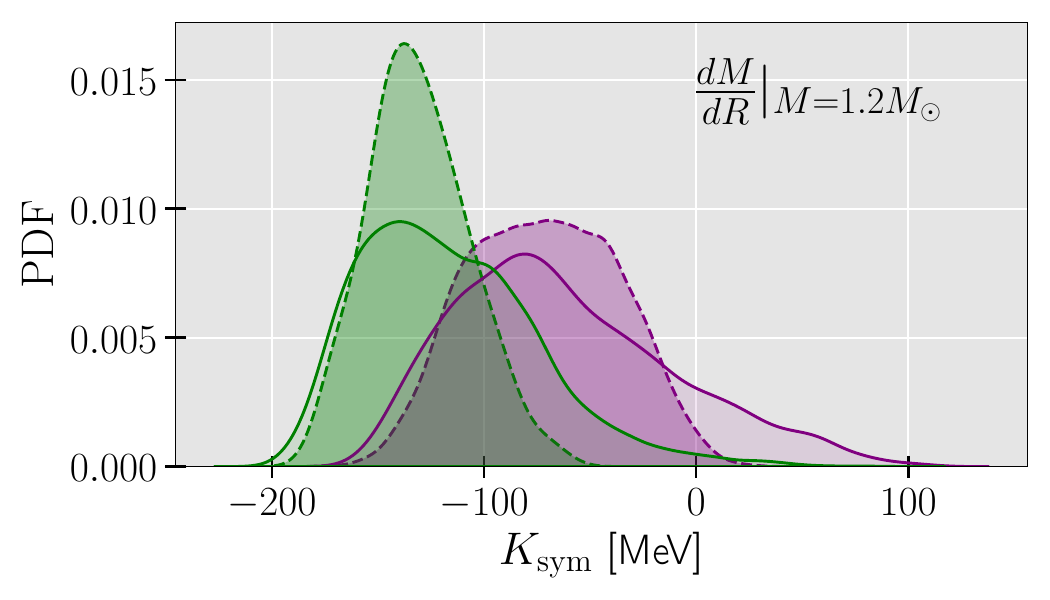}
    \includegraphics[width=0.32\linewidth]{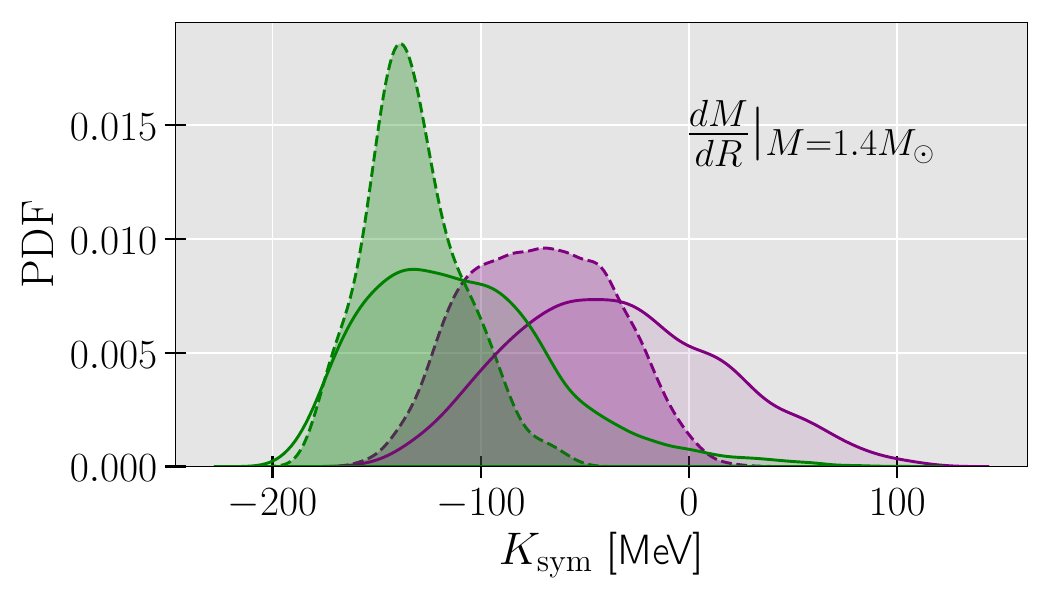}
    \includegraphics[width=0.32\linewidth]{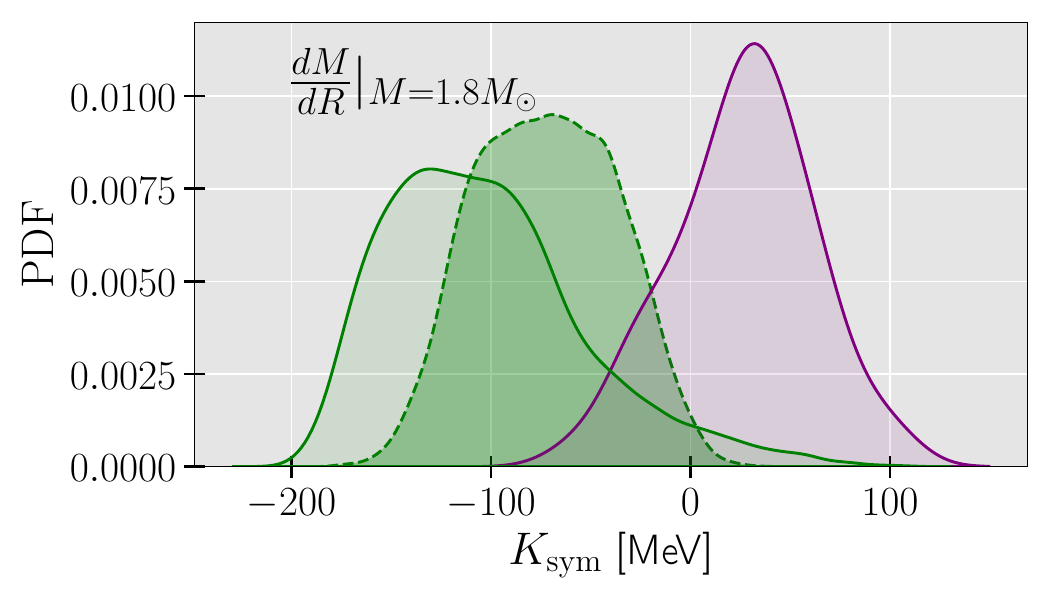}
    \caption{The PDFs of the $dM/dR|_{M}$ sign (positive/negative) at specific $M$ values for nucleonic (solid) and hyperonic (dashed) datasets. Purple indicates positive $dM/dR|_{M}$ values while green indicates negative values. The number of  hyperonic EoS satisfying positive/negative ($+/-$) $dM/dR|_{M}$ is  summarized in Table \ref{tab:new}. Note that for $M=1.8\, M_\odot$  the purple dashed distribution is missing because there are no hyperonic EoS with a positive $dM/dR$ for this mass, see Table \ref{tab:new}.}
      
    \label{fig:3}
\end{figure*}

The PDFs of \( dM/dR|_M \) at \( M/M_{\odot} = 1.2, 1.4, \) and \( 1.8 \) for the three isovector properties, $J_{\rm sym}$, $L_{\rm sym}$, and $K_{\rm sym}$, are shown in Fig.~\ref{fig:3}. 
We indicate the number of EoS in each set in Table \ref{tab:new}. Note that the number of hyperonic EoS with negative slope is very small for \( M/M_{\odot} = 1.2, 1.4, \) and  {for \( M/M_{\odot} = 1.8\) all  hyperonic EoS  have a negative slope}. This allows us to conclude that if a negative slope is measured for a low mass star there is a high probability that it is a nucleonic star. On the other hand, about one fourth of the nucleonic stars with \( M/M_{\odot} = 1.2, \) may have a positive slope. This fraction decreases to less than 0.1 for nucleonic stars with \( M/M_{\odot} = 1.4, \). A positive slope at \( M/M_{\odot} = 1.2, \) is not conclusive but at \( M/M_{\odot} = 1.4, \) could indicate the possible presence of hyperons.
\begin{table}[!htb]
\caption{Number of EoS in the different sets used in Sec.~\ref{sec:results2} on Fig.~\ref{fig:3}.}
\label{tab:new}
\begin{tabular}{ccccccc}
\toprule
$M/M_\odot$ & \multicolumn{2}{c}{1.2} &\multicolumn{2}{c}{1.4}& \multicolumn{2}{c}{1.8} \\
$dM/dR|_M$&    -&+&-&+&-&+\\
\midrule
hyperonic &95&16,501&193&15,953&16,146&0\\
nucleonic & 13,025& 4,511&16,023&1,514&17,537 &47\\
\bottomrule
\end{tabular}
\end{table}
A negative value of \( dM/dR|_M \) (green) at low and moderate NS masses constrains the symmetry energy (\( J_{\rm sym} \)), with the hyperonic set (dashed line) showing values larger of about 3-4 MeV than the nucleonic set (solid line). 
For \( dM/dR|_M > 0 \) (purple), the distribution $J_{\rm sym}$ remains quite similar throughout the mass range, the hyperonic case showing higher values, but with differences below 1 MeV, for the two smaller masses. Regarding the slope of the symmetry energy (\( L_{\rm sym} \)) for \( dM/dR|_M < 0 \) (green), the distributions differ between scenarios: the nucleonic case peaks at significantly lower values (\( L_{\rm sym} \sim 28 \) MeV), while the hyperonic case peaks at higher values (\( L_{\rm sym} \sim 50 \) MeV). 
However, the hyperonic sets with negative slope have a very small number of EoS.

Finally, the curvature of the symmetry energy (\( K_{\rm sym} \)) shows a clear separation depending on the sign of \( dM/dR|_M \). Although it remains difficult to distinguish between nucleonic and hyperonic scenarios, a negative \( dM/dR|_M \) (green) would suggest much lower \( K_{\rm sym} \) values compared to a positive \( dM/dR|_M \) (purple). The nucleonic cases allow for a wider distribution for \( K_{\rm sym} \), in particular for the largest mass considered $1.8\, M_\odot$ and \( dM/dR|_M >0\), this property peaks at positive values, unlike all other cases, and allows values above 100 MeV. However, this is a set with only 47 EoS. We may conclude that small values of \( K_{\rm sym} \), in particular, below $-100$ MeV are associated with a higher probability with \( dM/dR|_M < 0 \)  stars.

\section{\label{sec:loglikelihood} Scenarios Likelihood }
To quantitatively assess the agreement between each EoS and observational data, we calculate likelihoods based on pulsar and gravitational wave measurements (see \cite{Lackey:2014fwa,Raaijmakers:2019dks,Landry:2020vaw,Biswas:2020puz} for similar approaches).
 Specifically, we consider pulsar observations of PSR J0030+0451 \cite{Riley_2019,Miller19} and PSR J0740+6620 \cite{Riley2021,Miller2021} by NICER x-ray timing instrument, complemented in the case of PSR J0740+6620 by XMM-Newton event data, as well as the gravitational wave event GW170817 \cite{LIGOScientific:2018cki} detected by the LIGO-Virgo collaboration.

Herein, we are interested in determining the likelihood of each scenario and will not apply astrophysical cut-off constraints as we did in last sections, i.e., the radii inequality conditions for PSR J0740+6620, PSR J0030+0451, and $\Lambda(1.4M_\odot) < 720$ from GW170817 event. Instead, we are going to use the full dataset of \cite{Malik:2023mnx}, i.e. 17,828 nucleonic EoS and 18,756 hyperonic EoS, which is only limited by the properties of nuclear matter and $M_{\rm max}>2.0M_{\odot}$.   %\red{?? in each set how many (which fraction)  $dM/dR < 0$? } 
The nucleonic set includes 11,707 EoS that satisfy $dM/dR < 0$ and 6,121 obeying $dM/dR \not < 0$, while the hyperonic set consists of 79 EoS with $dM/dR < 0$ and 18,677 EoS which $dM/dR \not < 0$.

For a NS binary with component masses \( M_1 \) and \( M_2 \) (with $M_1\geq M_2$) and corresponding tidal deformabilities \( \Lambda_1 \) and \( \Lambda_2 \), the marginalized likelihood of a GW event for a given EoS is 
\begin{align}
\mathcal{L}^{\rm GW} &= P(d_{\mathrm{GW}}|\mathrm{EoS}) = \int_{M_l}^{M_u} \mathrm{d}M_1 \int_{M_l}^{M_1} \mathrm{d}M_2 \, P(M_1, M_2 | \mathrm{EoS}) \nonumber\\
 &\times P(d_{\mathrm{GW}} | M_1, M_2, \Lambda_1(M_1, \mathrm{EoS}), \Lambda_2(M_2, \mathrm{EoS})),
\end{align}
where \( P(d_{\mathrm{GW}} | M_1, M_2, \Lambda_1, \Lambda_2) \)
denotes the probability of the GW data given the masses and tidal deformabilities, and \( P(M_1, M_2 | \mathrm{EoS}) \) represents the prior on the component masses. We adopt a uniform prior,
\begin{equation}
P(M | \mathrm{EoS}) = 
\begin{cases}
\displaystyle\frac{1}{M_u - M_l}, & M_l \leq M \leq M_u, \\
0, & \text{otherwise},
\end{cases}
\end{equation}
where \( M_l = 1M_{\odot} \) and \( M_u = M_{\rm max} \), the maximum mass supported by the EoS.

{For the likelihood function $P(d_{\mathrm{GW}} \,|\, M_1, M_2, \Lambda_1, \Lambda_2)$, we used the joint marginalized posterior distribution $P(M_1, M_2, \Lambda_1, \Lambda_2 \,|\, d_{\mathrm{GW}})$ inferred by the LIGO/Virgo Collaboration \cite{LIGOScientific:2018cki}. Since the GW170817 analysis employed flat priors on the component masses and tidal deformabilities (see \cite{Raaijmakers:2019dks}), the posterior is proportional to the likelihood up to an overall normalization constant. We applied kernel density estimation (KDE) to the publicly available posterior samples of the GW170817 event \cite{LIGOScientific:2018cki} to obtain a smooth likelihood function. 
This approach directly employs the EoS-predicted component tidal deformabilities \( (\Lambda_1, \Lambda_2) \) at given masses, which are the fundamental quantities derived from the EoS, rather than the mass-weighted tidal deformability \( \tilde{\Lambda} \) primarily inferred from GW observations.
}\\

The NICER x-ray data for each pulsar is provided as posterior samples in the mass–radius plane. The marginalized likelihood associated with the NICER observations for a given EoS is
\begin{align}
\mathcal{L}^{\rm PSR} &= P(d_{\rm x-ray}|\mathrm{EoS}) \\
&= \int_{M_l}^{M_u} \mathrm{d}M \, P(M|\mathrm{EoS}) \, P(d_{\rm x-ray}|M, R(M, \mathrm{EoS})), \nonumber
\end{align}
where \( P(d_{\rm x-ray}|M, R) \) denotes the probability of the observed x-ray data given a specific neutron star mass \( M \) and its corresponding radius \( R \). For a given EoS, the radius \( R \) is uniquely determined by the mass \( M \), i.e., \( R(M, \mathrm{EoS}) \). The term $P(M|\mathrm{EoS})$ represents the prior on the individual pulsar mass for a given EoS, consistent with the prior defined for the component masses in the gravitational wave likelihood.

We compute \( P(d_{\rm x-ray}|M, R) \) using kernel density estimation applied to the publicly available \( M-R \) posterior samples from PSR J0030+0451 \cite{{Riley_2019,Miller19}} and PSR J0740+6620 \cite{Riley2021,Miller2021}. {Like for the GW likelihood function, the likelihood function \( P(d_{\rm x-ray}|M, R) \) was matched to the posterior distribution functions due to the flat priors used by the different analysis (see \cite{Raaijmakers:2019dks}).}
For each pulsar, the contributions from the two teams (Miller et al. and Riley et al.) are combined by averaging their respective log-likelihoods. Finally, the combined log-likelihoods from PSR J0030+0451 and PSR J0740+6620 are summed, under the assumption that these observations are independent.\\ 

\begin{figure*}[!htb]
    \centering
    \includegraphics[width=0.329\linewidth]{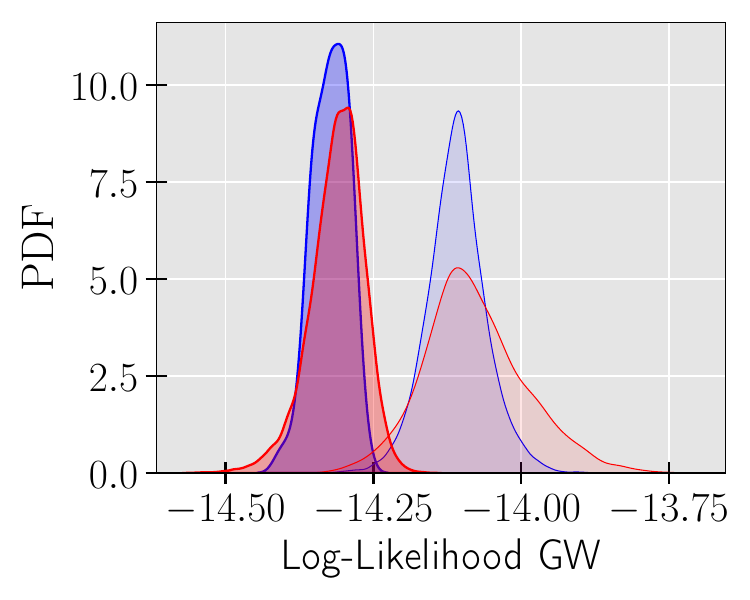}
    \includegraphics[width=0.329\linewidth]{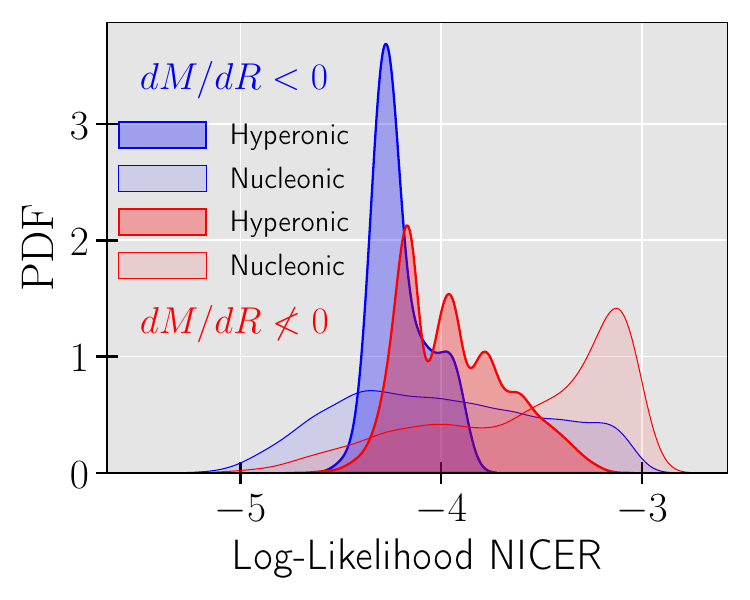}
    \includegraphics[width=0.329\linewidth]{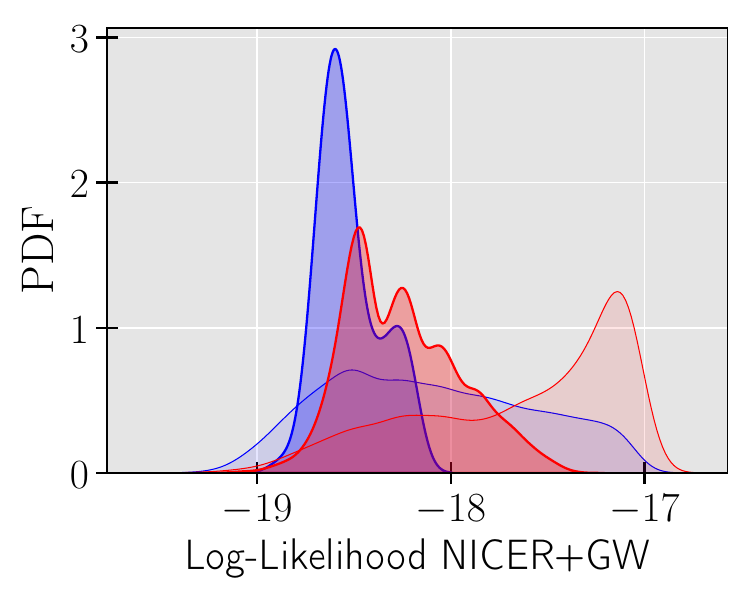}
    \caption{Log-likelihood (see Eq.~\ref{eq:log_lik}) of the hyperonic (dark colors) and hadronic (light colors) sets for   $dM/dR < 0$ (blue) and $dM/dR \not < 0$ (red). 
    The $\log \mathcal{L}^{\rm GW}$ is displayed on the left, $\log \mathcal{L}^{\rm NICER}$ on the center, while the total log-likelihood $\log \mathcal{L}$ is on the right.}
    \label{fig:4}
\end{figure*}

Assuming the NICER (complemented with the XMM-Newton data in the case of PSR J0740+6620)  and GW datasets are independent, the total likelihood is given by the product
\begin{equation} 
\log \mathcal{L} = \log \mathcal{L}^{\rm GW} + \log \mathcal{L}^{\rm {PSR}}.
\label{eq:log_lik}
\end{equation}
The individual and total log-likelihoods are shown in Fig.~\ref{fig:4}.  
We find that the GW170817 event is better explained within the nucleonic scenario (left panel), the corresponding EoS is more consistent  with the observed data as indicated by the higher likelihood.  Additionally, we observe that the nucleonic distribution becomes slightly broader in the  case of a non-zero derivative of the mass-radius relation \( dM/dR \not< 0 \) (red), corresponding to the presence of back-bending in the \( M(R) \) relation.  We expect a lower value of \( \log \mathcal{L}^{\rm GW} \) for the hyperonic scenario, since it predicts systematically larger  value of \( \Lambda(1.4 M_{\odot}) \) compared to the nucleonic case. This leads to a lower agreement with the GW170817 constraints. The individual log-likelihoods for the NICER observations show a much wider spread of PDFs for  the nucleonic EoS. This is  also to  be expected given that the spread of the mass-radius distributions shown  in Fig. \ref{fig:1} are much wider for the nucleonic EoS  and contain the hyperonic ones.\\ 

The information in Fig.~\ref{fig:4} is supplemented by the Bayes factors presented in Table \ref{tab:2} for each composition (nucleonic or hyperonic) and hypothesis regarding the slope behavior $dM/dR$. The Bayes factor $B_{ij} = Z_i / Z_j$ is the ratio of evidences for hypotheses $\mathcal{H}_i$ and $\mathcal{H}_j$, where the evidences $Z_i=p(d_{\rm x-ray},d_{\rm GW}  | \mathcal{H}_i)$ are given by 
\begin{equation}
Z_i = \frac{1}{N} \sum_{j=1}^N p(d_{\rm x-ray},d_{\rm GW}  | \text{EoS}_j, \mathcal{H}_i),
\end{equation}
where $p(d_{\rm x-ray},d_{\rm GW}  | \text{EoS}_j, \mathcal{H}_i)$ combines the likelihoods of GW and NICER data, assuming their independence given an EoS.  A Bayes factor $B_{ij} > 1$ indicates that $\mathcal{H}_i$ is more compatible with $\{d_{\rm x-ray},d_{\rm GW}\}$ than $\mathcal{H}_j$, while $B_{ij} < 1$ favors $\mathcal{H}_j$. These Bayes factors are critical for model comparison, providing a quantitative measure to assess which hypothesis best explains the observed GW and x-ray data, thus guiding conclusions about competing astrophysical models, such as neutron star EoS. In summary, the Bayes factors suggest that the astrophysical constraints favor: i) nucleonic matter with $dM/dR \not< 0$ over $dM/dR < 0$ ($B_{12} < 1$); ii) nucleonic matter, regardless of the slope, over hyperonic matter ($B_{ij} > 1$, for $i=1,2$ and $j=3,4$).

\begin{table}[!htb]
\caption{Bayes factors comparing the four hypotheses for nucleonic and hyperonic matter scenarios. Hypotheses are defined as: nucleonic with $dM/dR < 0$ ($\mathcal{H}_1$), nucleonic with $dM/dR \not < 0$ ($\mathcal{H}_2$), hyperonic with $dM/dR < 0$ ($\mathcal{H}_3$), and hyperonic with $dM/dR \not < 0$ ($\mathcal{H}_4$). Each entry represents the Bayes factor \( B_{ij} = Z_i / Z_j \), where \( Z_i \) is the evidence for hypothesis $\mathcal{H}_i$, and the labels $i$ and $j$ correspond, respectively to lines and columns. The lower triangule values are given by $B_{ji}=1/B_{ij}$ (see text for details).}
\label{tab:2}
\centering
\begin{tabular}{l *{4}{S[table-format=2.4, round-mode=places, round-precision=4]}}
\toprule
& { $\mathcal{H}_1$} & {$\mathcal{H}_2$} & {$\mathcal{H}_3$} & {$\mathcal{H}_4$} \\
& { Nucleonic} & {Nucleonic} & {Hyperonic} & {Hyperonic} \\
  & { $dM/dR < 0$} & {$dM/dR \not < 0$} & { $dM/dR < 0$} & {$dM/dR \not < 0$} \\
\midrule
$\mathcal{H}_1$ & {1} & {0.643} & {1.722} & {1.270} \\
$\mathcal{H}_2$ & {-} & {1} & {2.680} & {1.976} \\
$\mathcal{H}_3$ & {-} & {-} & {1} & {0.737} \\
$\mathcal{H}_4$ & {-} & {-} & {-} & {1} \\
\bottomrule
\end{tabular}
\end{table}

\section{\label{sec:conclusions}Conclusions}

In the present study we have analyzed the possible information that the derivative of the mass radius curve may give concerning the composition of the neutron star and its EoS properties. We have considered two sets of about 17k-18k EoS obtained within a Bayesian inference calculation to determine the couplings of a relativistic mean field description of nucleonic and hyperonic matter \cite{Malik:2022zol}. The RMF model considered has been discussed in \cite{Mueller1996,Horowitz:2000xj,Todd-Rutel2005}, and includes nonlinear mesonic terms, both self-interacting and mixed terms. The nuclear matter properties of the two sets are summarized in Table \ref{tab:3}.

The two sets were divided into two subsets each, defined by the slope of the complete mass-radius curve, from the 1$M_\odot$ mass to the maximum mass, filtered by conditions $dM/dR < 0$ and $dM/dR \not < 0$. Whereas in the nucleonic set about one third satisfied $dM/dR < 0$, this property was satisfied by less than 100 hyperonic EoS. 

Our results are summarized as follows: i) a negative slope along the entire mass-radius curve indicates a large probability of the absence of hyperons. This conclusion can already be drawn if the slope is negative for low mass stars, with a mass below 1.2$M_\odot$ or even 1.4$M_\odot$. While a positive slope for low-mass stars does not necessarily indicate the presence of hyperons, it is not expected that the mass-radius curve will still have a positive slope at 1.8$M_\odot$; ii) larger maximum masses are obtained for nucleonic models with $dM/dR \not < 0$, the maximum mass obtained being $\sim 0.2M_\odot$ larger at {90}\%CI; iii) in the average nucleonic EoS with $dM/dR \not < 0$ predict slightly larger radius for 1.4$M_\odot$ stars, about 150 m, although differences of 300 m are possible. Similarly, the tidal deformability appears to be larger for these stars; iv) the mass-radius curves of nucleonic stars with $dM/dR < 0$ correspond to EoS with lower maximum values of pressure, sound speed, and trace anomaly. 

Our main conclusions can be summarized as: within our  covariant density functional description there is a clear indication that mass-radius curves with a negative slope for all masses between 1$M_\odot$ and the maximum mass cannot result from an EoS that also contains non-nucleonic degrees of freedom (hyperons) and a positive slope at 1.4$M_\odot$ is a strong indication that the EoS contains  non-nucleonic degrees of freedom.

We have also studied the nuclear matter properties of stars with a positive or negative slope at masses 1.2, 1.4, and 1.8$M_\odot$. The few hyperonic EoS with $dM/dR < 0$ for 1.2 and 1.4$M_\odot$ stars must have a rather high symmetry energy at saturation, about 3-4 MeV higher than all other scenarios, nucleonic with $dM/dR < 0$ or $dM/dR \not < 0$ or hyperonic with $dM/dR \not < 0$. 
For 1.2 and 1.4$M_\odot$ stars with a positive slope in their mass-radius curves, there is little difference between the nuclear matter parameters of nucleonic and hyperonic stars.  The property that most distinguishes the  nucleonic scenarios with positive slope is the curvature of the symmetry energy, which can take positive values above 100 MeV. For the few  1.8$M_\odot$ nucleonic stars with a positive slope of the mass-radius curve, the median of the curvature of the symmetry energy takes positive values around 50 MeV.

The likelihood distributions and respective Bayes factors suggest that nucleonic EoS are more probable than the hyperonic ones, given the GW170817 and NICER observation constraints for PSR J0030+0451 and PSR J0740+6620. From the two nucleonic sets of EoS, the one with $dM/dR \not < 0$ has the highest probability. These conclusions are in line with the ones drawn in \cite{Carvalho:2024bxv}, where using a neural network classification model for detecting the presence of hyperonic degrees of freedom, the authors have found from a set of observational data 
that the presence of hyperons inside NSs was not favored. In \cite{Huang:2024rvj} the authors also analyse whether measurements of properties neutron star mass and radius could identify the presence of hyperons inside neutrons stars. While current observations do not distinguish between the two scenarios, it was shown using simulated data that it would be possible to identify the presence of hyperons. 

The present study has demonstrated the potential power of analysing the slope of mass-radius curves. We expect that, with tighter constraints, more conclusive compositional information can be obtained in the future.  The onset of other non-nucleonic degrees of freedom, such as delta-baryons or kaon condensates, could have effects similar to those of the onset of hyperons. Information on the mass-radius slope may also help the interpretation of results obtained using agnostic descriptions. Interestingly, of the six 'golden' agnostic EoS chosen from the 68\% CI in \cite{Ecker:2024uqv} three of them have a positive slope at 1.8$M_\odot$, which seems to indicate matter with properties different from the present study.  However, the choice in \cite{Ecker:2024uqv} also reflects the lower probability that a mass radius curve has negative slope in its full range as obtained in our study.  

\section*{ACKNOWLEDGMENTS} 
We thank Tuhin Malik for kindly providing the codes used to calculate the likelihood values. This work was partially supported by national funds from FCT (Fundação para a Ciência e a Tecnologia, I.P, Portugal) under the projects 2022.06460.PTDC with the  DOI identifier 10.54499/2022.06460.PTDC, and 
UIDB/04564/2020 and UIDP/04564/2020, with DOI identifiers 10.54499/UIDB/04564/2020 and 10.54499/UIDP/04564/2020, respectively.

\onecolumngrid
\appendix

\section{Datasets statistical summary\label{ap:stats}}
Table \ref{tab:1} shows some statistics regarding the nuclear matter parameters of each composition scenario (nucleonic/hyperonic) and $dM/dR$ behavior ($<0$ or $\not <0$).

\begin{table*}[!htb]
\caption{The 5\%, 50\%, and 95\% quantiles for the nuclear matter properties of the nucleonic and hyperonic datasets  given in \cite{malik_2023_7854112} .  All quantities are given in MeV.}
\label{tab:3}
\centering
%\resizebox{\columnwidth}{!}{%
\begin{tabular}{c ccccccc c ccccccc }
\toprule
&  \multicolumn{7}{c}{Nucleonic} & &  \multicolumn{7}{c}{Hyperonic} \\
\cline{2-8} \cline{10-16} 
& \multicolumn{3}{c}{$dM/dR<0$}  & &\multicolumn{3}{c}{$dM/dR\nless0$} & &\multicolumn{3}{c}{$dM/dR<0$}  & &\multicolumn{3}{c}{$dM/dR\nless0$}  \\
\cline{2-4} \cline{6-8} \cline{10-12} \cline{14-16} 
& 5\% & 50\% & 95\% & & 5\% & 50\% & 95\% & & 5\% & 50\% & 95\% & & 5\% & 50\% & 95\%\\ 
 \midrule
$E_0$  &-16.41&-16.09&-15.76 & &-16.46&-16.11&-15.77 & &-16.32&-16.01&-15.63 & &-16.41&-16.09&-15.77    \\
$K_0$  &217&252&289 & &205&260&307 & &262&289&335 & &269&293&337   \\
$Q_0$  &-523&-452&-349 & &-502&-399&-123 & &-281&-168&-43 & &-251&-132&44    \\
$Z_0$ &490&1925&4056 & &-104&2015&6858 & &-991&1869&2861 & &-313&2133&3023    \\
$J_{\text{sym}}$   &30.3&32.4&34.6 & &28.3&30.8&33.3 & &33.4&35.1&36.7 & &29.3&31.9&34.4    \\
$L_{\text{sym}}$   &21.9&37.0&57.7 & &25.3&37.9&58.1 & &35.3&45.4&56.2 & &30.4&40.5&56.5    \\
$K_{\text{sym}}$   &-175&-124&-43 & &-141&-79&28 & &-170&-134&-96 & &-127&-72&-16    \\
$Q_{\text{sym}}$   &685&1450&1738 & &507&1262&1524 & &1146&1570&1848 & &934&1332&1614    \\
$Z_{\text{sym}}$   &-19477&-12181&-197 & &-17989&-13138&-1339 & &-19113&-11414&-4020 & &-19062&-13797&-3871    \\
\bottomrule
\end{tabular}
%}
\end{table*}

\bibliographystyle{apsrev4-1}
%\bibliography{biblio}
\providecommand{\noopsort}[1]{}\providecommand{\singleletter}[1]{#1}%

\end{document}